\documentclass[letterpaper]{article} 
\usepackage{times}  
\usepackage{helvet}  
\usepackage{courier}  
\usepackage{url}  
\usepackage{graphicx}  
\usepackage[ruled,linesnumbered]{algorithm2e}
\usepackage[numbers,square]{natbib}	

\usepackage{tikz}
\usetikzlibrary{shapes.arrows,chains}
\usetikzlibrary{fit}
\usetikzlibrary{shapes.geometric}
\usetikzlibrary{arrows,automata}
\usetikzlibrary{shapes,positioning}

\tikzstyle{DecisionNodeMin} = [ellipse, black, font=\bfseries, draw=black, align=center, minimum height=.5cm,minimum width=.5cm, inner sep=0pt, text centered]
\tikzstyle{DecisionNodeMax} = [rectangle, black, draw=black, align=center, minimum height=.4cm,minimum width=.4cm, inner sep=0pt, text centered]
\tikzstyle{Leaf} = [regular polygon,regular polygon sides=5, black, draw=black, minimum height=.5cm,minimum width=.5cm, align=center, inner sep=0pt, text centered]

\tikzset{
  treenode/.style = {align=center, inner sep=0pt, text centered,
    font=\sffamily},
  arn_x/.style = {treenode, circle, draw=black, fill=black,
    minimum width=0.2em, minimum height=0.2em},
    arn_y/.style = {treenode, rectangle, draw=black, fill=black,
    minimum width=0.2em, minimum height=0.2em}
}

\tikzset{
  invisible/.style={opacity=0},
  visible on/.style={alt=#1{}{invisible}},
  alt/.code args={<#1>#2#3}{%
    \alt<#1>{\pgfkeysalso{#2}}{\pgfkeysalso{#3}} 
  },
}

\usepackage{amssymb}
\usepackage{amsmath}
\usepackage{amsthm}

\tikzstyle{AllNode} = [circ,minimum height=.7cm,minimum width=.7cm, black, draw=black, align=center, inner sep=1pt, text centered, solid]
\tikzstyle{ExNode} = [regular polygon,regular polygon sides=5,minimum height=.7cm,minimum width=.7cm, black, draw=black, align=center, inner sep=1pt, text centered, solid]

\newtheorem{my}{Proof}

\newtheorem{mydef}{Definition}
\newtheorem{myTheo}{Theorem}

\newtheorem{myex}{Example}

\providecommand{\keywords}[1]{\textbf{\textit{Keywords: }} #1}

\title{Game Tree Search in a Robust Multistage Optimization Framework: Exploiting Pruning Mechanisms}
\author{
Michael Hartisch and Ulf Lorenz \\ University of Siegen, Siegen, Germany\\ \{michael.hartisch, ulf.lorenz\}@uni-siegen.de
}
\date{}
\begin{document}
%

\maketitle
\begin{abstract}
We investigate pruning in search trees of so-called quantified integer linear programs (QIPs). QIPs consist of a set of linear inequalities and a minimax objective function, where some variables are existentially and others are universally quantified. They can be interpreted as two-person zero-sum games between an existential and a universal player on the one hand, or multistage optimization problems under uncertainty on the other hand. Solutions are so-called winning strategies for the existential player that specify how to react on moves of the universal player - i.e. certain assignments of universally quantified variables - to certainly win the game.

QIPs can be solved with the help of game tree search that is enhanced with non-chronological back-jumping. We develop and theoretically substantiate pruning techniques based upon (algebraic) properties similar to pruning mechanisms known from linear programming and quantified boolean formulas. The presented Strategic Copy-Pruning mechanism allows to \textit{implicitly} deduce the existence of a strategy in linear time (by static examination of the QIP-matrix) without explicitly traversing the strategy itself. We show that the implementation of our findings can massively speed up the search process.
\end{abstract}
\section{Introduction}
Mixed-integer linear programming (MIP) \cite{Schrijver} is the state-of-the-art technique for computer aided optimization of real world problems. Nowadays, commercial top solvers  are able to solve large MIPs of practical size, but companies observe an increasing danger of disruptions, which prevent them from acting as planned. One reason is that input
data is often assumed to be deterministic and exactly known
when decisions have to be made, but in reality they are often afflicted with some kinds of uncertainties. Examples are flight and travel times, throughput times or
arrival times of externally produced goods. Thus, there is a need for planning and
deciding under uncertainty. Uncertainty, however, often pushes the complexity
of problems that are in the complexity class P or NP, to PSPACE \cite{Papadimitriou}. Therefore, NP-complete integer programs are not appropriate for such problems. Prominent solution paradigms for optimization under uncertainty are
Stochastic Programming \cite{Birge}, Robust Optimization \cite{Ben-Tal,Liebchen2009,Goerigk2016}, Dynamic Programming
\cite{Bellman}, Sampling \cite{GuptaA} and of course POMDP \cite{nguyen2017collective}. 
Relatively unexplored are the abilities of linear programming extensions for
PSPACE-complete problems. In the early 2000s the idea of universally quantified variables, as they are used in quantified constraint satisfaction problems \cite{Gerber}, was picked up again \cite{Subramani}, coining the term quantified integer program (QIP). Quantified integer programming  is a direct, very formal extension of integer linear programming (IP), making QIPs applicable in a very natural way \cite{ESA11,ederer2016multistage}. They allow robust multistage optimization extending the two/three-stage approach of Robust Optimization \cite{Ben-Tal}. Multistage models in contrast to two/three-stage models allow more precise planning strategies as uncertain events typically do not occur all at the same time (delay in timetables, changed cost estimate for edges in a graph).

  A solution of a QIP is a strategy -- in the game tree search sense \cite{Pijls}, see Definition \ref{Def_Strat} -- for assigning existentially quantified variables such that some linear constraint system is fulfilled. By adding a minimax objective function the aim is to find the best strategy \cite{Euro}. As not unusual in the context of optimization under uncertainty \cite{Ben-Tal,Bertsimas} a polyhedral uncertainty set can be used \cite{CG16}. There are two different ways known how to tackle a QIP: On the one hand the so-called deterministic equivalent program can be built, similar to the ones known from stochastic programming \cite{Wets}, and solved using standard integer programming solvers. On the other hand the more direct approach is to conduct a game tree search \cite{Allis,Schaeffer,Feldmann,GoogleGo}. We are interested in utilizing game solving techniques in combination with linear programming techniques.  Recently a solver for quantified mixed integer programs was made  available as open source. This solver combines techniques known from game tree search, linear programming and (quantified) boolean formula \cite{YasolACG17}. 

An optimization task is often split up into two parts: finding the optimal solution itself and proving that no better solution can exist. It turned out that applying backjumping techniques as utilized by QBF-solvers \cite{Zhang} and cutting planes as commonly used in integer programming \cite{Cuts} are also highly beneficial for QIPs in order to assess that no (better) strategy can exist in certain subtrees. This subtask even becomes simpler with an increasing number of universally quantified variables. However, finding a solution, which we call a winning strategy, proved to be more difficult. At first glance, it seems that the exponential number of leaves belonging to a strategy must be traversed explicitly. This is certainly true in the worst-case. However, as practical game trees are often structured irregularly,  typically there are ``difficult'' parts of a game tree where a very deliberated substrategy must be found but also other parts that are really easy to master. In this paper we present a procedure, called strategic copy-pruning (SCP), that is capable of recognizing such easily-masterable subtrees which makes it possible to \textit{implicitly} deduce the existence of a winning strategy therein.  In experiments, this SCP often allows to conclude the existence of a winning strategy with a linear number of algebraic operations. In particular, in those cases it is not necessary to examine an exponential number of leaves.

The effect of SCP is reinforced if the sequence of variable assignments predicted as optimal by minimax for both sides, called the principal variation \cite{Minimax}, is traversed in an early stage of the tree search. Detecting and verifying this particular variable assignment is essential in order to obtain the objective value. This, of course, is not as easy as it sounds, but having reasonable knowledge of which universal variable assignments are particularly vicious can massively boost the search process. Several heuristics exist to analyze and find such promising moves in a game tree search environment \cite{Killer,Schaeffer,Plaat:1,WinandsHistory}.



Of course there are  publications specifically dealing with pruning and backjumping techniques both in the area of game tree search \cite{WinandsPruning,KNUTH1975293,Hauk} and quantified boolean formula (QBF) \cite{Qube,Cadoli2002} as well. Moreover there is Kawano's simulation \cite{Kawano}, sss* \cite{Stockman}, MTD(f) \cite{Plaat:1,Plaat:1996} and (nega)scout \cite{Reinefeld}. 
However, none of the above cover what we do here.

The paper is organized as follows: First basic definitions and notations regarding QIPs are presented. Then two pruning techniques for the QIP game tree search are introduced and examined theoretically: First, the well known  monotonicity \cite{Cadoli2002} of variables is recaptured. Second, as our main result, we  derive from already found strategies the existence of winning strategies in other branches. This happens in a way such that these branches do not need to be investigated explicitly. Finally the conducted experiments are presented.

\section{Preliminaries: Basics of Quantified Integer Programming}\label{Sec_QIP}
Let $n\in \mathbb{N}$ be the  number of variables and $x=(x_1, \ldots, x_n)^\top	 \in \mathbb{Z}^n$ a vector
 of variables.\footnote{$\mathbb{Z}$, $\mathbb{N}$ and $\mathbb{Q}$ denote the set of integers, natural numbers, and rational numbers, respectively.}
For each variable $x_j$ its domain $\mathcal{L}_j$ with $l_j, u_j \in \mathbb{Z}$, $l_j \leq u_j$, $1 \leq j \leq n$, is given by
$\mathcal{L}_j =\{y\in \mathbb{Z} \mid    l_j \leq y \leq u_j \}$.
The domain of the entire variable vector is described by $\mathcal{L} =\{y \in \mathbb{Z}^n \mid \forall j \in \{1,\ldots,n\}: y_j \in \mathcal{L}_j\}$, i.e. each variable must obey its domain.
Let $Q \in \{\exists, \forall\}^n$ denote the vector of quantifiers. We call $\mathcal{E}= \{j \in \{1,\ldots, n\} \mid Q_j=\exists \}$ the set of existential variables and $\mathcal{A}= \{j \in \{1,\ldots, n\} \mid Q_j=\forall \}$ the set of universal variables. Further, each maximal consecutive subsequence in $Q$ consisting of identical quantifiers is called \textit{quantifier block} with $B_i \subseteq \{1,\ldots,n\}$ denoting the $i$-th block.  Let $\beta \in \mathbb{N}$, $\beta \leq n$, denote the number of blocks and thus $\beta-1$ is the number of quantifier changes. The variable vector of variable block $B_i$ will be referred to as $x^{(i)}$.

\begin{mydef}[Quantified Integer Linear Program (QIP)] \label{Def_QIP}~\\
Let $A \in \mathbb{Q}^{m \times n}$ and $b \in \mathbb{Q}^{m}$ for $m \in \mathbb{N}$ and let $\mathcal{L}$ and $Q$ be given as described above.  Let $c \in \mathbb{Q}^n$ be the vector of objective coefficients and let $c^{(i)}$ denote the vector of coefficients belonging to block $B_i$. Let the term $Q \circ x \in \mathcal{L}$ with the component wise binding operator $\circ$ denote the \emph{quantification vector} $(Q_1 x_1 \in \mathcal{L}_1, \ldots, Q_n x_n \in\mathcal{L}_n)$ such that every quantifier $Q_j$ binds the variables  $x_j$  to its domain $\mathcal{L}_j$. We call $(A,b,c,\mathcal{L},Q)$ with
\begin{align*}
&z={\min\limits_{B_1}\left( c^{(1)}x^{(1)} +\max\limits_{B_2} \left(c^{(2)}x^{(2)} + \ldots \min\limits_{B_\beta} c^{(\beta)}x^{(\beta)} \right)\right)} \\
& \text{s.t.\ } Q \circ x \in \mathcal{L}:\ A x \leq b \tag{*}
\label{EquationQIP}
\end{align*}

	a QIP with objective function (for a minimizing existential player). 
\end{mydef}
In the following, we will only consider binary QIPs, i.e. $l_j=0$ and $u_j=1$ for all $j\in \{1,\ldots,n\}$. This requirement, however, does not constitute a restriction as any QIP instance can be  converted artificially  through binarization.

A QIP instance can be interpreted as a two-person zero-sum game between an \emph{existential player} setting the existentially quantified variables and a \emph{universal player} setting the universally quantified variables with payoff $z$. The variables are set in consecutive order according to the variable sequence. Consequently, we say that a player makes the move $x_k = y$ if she fixes the variable  $x_k$ to $y \in \mathcal{L}_k$. At each such move, the corresponding player knows the settings of $x_1, \ldots, x_{k-1}$ before taking her decision $x_k$.  Each fixed vector $x \in \mathcal{L}$, that is, when the existential player has fixed the existential variables and the universal player has fixed the universal variables, is called \emph{a game}.  If $x$ satisfies the linear constraint system $Ax \leq b$, the existential player pays $z=c^\top x$ to the universal player. If $x$ does not satisfy $A x \leq b$, we say \emph{the existential player loses} and the payoff will be $+\infty$. This is a small deviation from conventional zero-sum games but using\footnote{Since this is only a matter of interpretation the consequences are not discussed further.}  $\infty+(-\infty)=0$ also fits for zero-sum games.
The chronological order of the variable blocks given by $Q$ can be represented using a game tree.

\begin{mydef}[Game Tree]\label{Def_Tree}~\\
Let $G=(V,E,c)$ be the edge-labeled finite directed tree with a set of nodes $V=V_\exists \cup\ V_\forall \cup\ V_L $, a set of edges $E$ and a vector of edge labels $c \in \mathbb{Z}^{\vert E \vert}$. Each  inner level of the tree either consists of only nodes from \(V_\exists\) or only of nodes from \(V_\forall\), with the root node at level 0 being from \(V_\exists\). The leaf nodes (nodes without children) are from $V_L$. The $j$-th variable is represented by the inner nodes at depth $j-1$. Each edge connects a node in some level \(j\) to a node in level \(j+1\). Outgoing edges from a node in level $j$ represent moves from $\mathcal{L}_{j+1}$ of the player at the current node, the corresponding edge labels encode the variable assignments of the move. 
\end{mydef}

Thus, a path from the root to a leaf represents a game of the QIP and the sequence of edge labels encodes its moves and hence the assignment of the corresponding variables. The most relevant term in order to describe solutions are so-called strategies.
\begin{mydef}[(Existential) Strategy]\label{Def_Strat}~\\
	A \emph{strategy} (for the assignment of existential variables) $S = (V',E',c')$ is a subtree of a game tree $G=(V,E,c)$. Each node \( v_\exists \in V' \cap  V_\exists \) has exactly one child, and each node \( v_\forall \in  V' \cap V_\forall \) has as many children as in $G$, i.e. as many as there are values in the corresponding variable domain.
\end{mydef}
In the following, the word \textit{strategy} will always refer to an \textit{existential} strategy. Universal strategies can be defined similarly but are not needed in our context. A strategy is called a \emph{winning strategy} if all paths from the root node to a leaf represent a vector \(x\) such that \(A x \leq b \). A QIP is called \emph{feasible} if (\ref{EquationQIP}) is true (see Definition \ref{Def_QIP}), i.e. if a winning strategy for the assignment of existential variables exists.
If there is more than one winning strategy for the existential player, the objective function aims for a certain (the ``best'') one. The value of a strategy is given by its minimax value which is the maximum value at its leaves \cite{Pijls}. Hence, the objective value of a feasible QIP is the minimax value of the root node, i.e. the minimax value of the optimal winning strategy. Note that a leaf not fulfilling $A x \leq b$ can be represented by the value $+\infty$. The minimax value for any node is recursively defined as follows:
\begin{mydef}[Minimax Value]\label{Def_MinMax}~\\
Let $S = (V',E',c')$ be a subtree of a game tree $G=(V,E,c)$ of a QIP as in Definition \ref{Def_Tree}. Let $x_v$ denote the variable assignment corresponding to the leaf node $v \in V_L$ defined by the edge labels of the path from the root to $v$. For any node $v \in V'$ the minimax value $f(v)$ is recursively defined by
$$f(v)=\begin{cases} 
c^\top x_v & \text{ , if $v \in V_L \wedge A x_v \leq b$}\\
+\infty \hspace{-.3cm}& \text{ , if $v \in V_L \wedge A x_v \not\leq b$}\\
\min\{f(v') \mid (v,v') \in E'\}\hspace{-.4cm}& \text{ , if $v  \in V_\exists$}\\
\max\{f(v') \mid (v,v') \in E'\}\hspace{-.4cm}& \text{ , if $v  \in V_\forall$}\ .
\end{cases}$$
For $S=G$ the value $f(v)$ of any node $v \in V$ is the outcome if the remaining variables are assigned optimally starting from this node, i.e. the outcome of optimal play by both players, whereas $f(v)=+\infty$ implies that there exists no (existential) strategy to ensure $A x \leq b$.
\end{mydef}
Hence, the value of a strategy $S = (V',E',c')$ is the minimax value of its root node and is defined by the \emph{principal variation} (PV) \cite{Minimax}, i.e. the sequence of variable assignments being chosen during optimal play in $S$. From now on $f(v)$ will  refer to  the outcome of optimal play by both players  in the entire game tree $G$, i.e. Definition \ref{Def_MinMax} with $S=G$ is used.
\begin{myex} Let us consider a QIP with $n=4$ binary variables, $Q=(\exists,\forall,\exists,\forall)$, $c=(2,-2,-3,-2)$ and let the constraint system $Ax \leq b$ given by
\[
\begin{array}{crcrcrcrcl}
				& x_1 	& + 	& x_2 	&+&x_3		&  &&\leq	& 2  \\
				& -x_1 	& 		& 			&+&x_3		&-	& x_4 &\leq	& 0 	\\
				&  		& -	& x_2	&+&x_3		&-	& x_4& \leq	& 0 	\\
				& -x_1	& +	& x_2	&-&x_3		&+& x_4& \leq	& 1\ .
\end{array}
\]
The minimax value of the root node of the game tree is $2$ and the principal variation is given by $x_1=1$, $x_2=0$, $x_3=0$ and $x_4=0$. The minimax value of the inner node at level 1 resulting from setting $x_1=0$ has the minimax value $+\infty$, i.e. after setting $x_1=0$ there exists no winning strategy. 

\end{myex}


\section{Theoretical Analysis}\label{Sec_Theory}
A quantified integer program can be solved using its deterministic equivalent program \cite{Wets}, which is an integer program with exponentially increased size, or via the more direct approach: a game tree search. During such a game tree search we are interested in quickly evaluating or estimating the minimax value of nodes, i.e. we want to examine the optimal (existential)  strategy of the corresponding subtree. 
In order to speed up the search process, limiting the number of subtrees that need to be explored is extremely beneficial. Such pruning operations are applied in many search based algorithms, e.g. the alpha-beta algorithm \cite{KNUTH1975293}, branch-and-bound \cite{Nemhauser} and DPLL \cite{Zhang}. In the following, we will present two approaches that allow pruning in a QIP game tree search, and thus in a strategic optimization task. 
\\

In case of QIPs a rather simple argument exists such that certain variable assignments never need to be checked as they are worse than their counterparts. This concept of monotone variables is well known in the field of quantified boolean formulas \cite{Cadoli2002} and integer programming \cite{Nemhauser}. 
We shortly present the consequences for QIPs, before we deal with our main Theorem \ref{Theo_CopyPaste}.
\begin{mydef}[Monotone Variable]~\\
A variable $x_k$ of a QIP is called monotone if it occurs with only positive or only negative sign in the matrix and objective, i.e. if the entries of $A$ and $c$ belonging to $x_k$ ($A_{\star,k}$ and $c_k$) are either all non-negative or all non-positive\footnote{$A_{\star,k}$ denotes the $k$-th column and $A_{i,\star}$ the $i$-th row  of $A$.}.
\end{mydef}

\begin{myTheo} \label{Therorem_Mono}
Let $(A,b,c,\mathcal{L},Q)$ be a QIP and let variable $x_k$, $1\leq k \leq n$, be monotone with all non-negative entries. For any two leaves $v^{(0)}$ and $v^{(1)}$ of the game tree represented by the fixed variable vectors $\tilde{x}^{(0)}=(\tilde{x}_1,\ldots, \tilde{x}_{k-1},0,\tilde{x}_{k+1},\ldots,\tilde{x}_n)\in \mathcal{L}$ and $\tilde{x}^{(1)}=(\tilde{x}_1,\ldots, \tilde{x}_{k-1},1,\tilde{x}_{k+1},\ldots,\tilde{x}_n) \in \mathcal{L}$, respectively, it is
$f(v^{(0)})\leq f(v^{(1)})$.
\end{myTheo}

\begin{my}
If $A\tilde{x}^{(0)}\not\leq b$ it is $f(v^{(0)})=+\infty$. Hence, some constraints $i \in \{1,\ldots,m\}$ exists with $b_i< A_{i,\star}\tilde{x}^{(0)}$. Due to the monotonicity of variable $k$ it is $A_{i,\star}\tilde{x}^{(0)}\leq  A_{i,\star}\tilde{x}^{(1)}$ and hence $f(v^{(1)})=+\infty$, i.e. $f(v^{(0)})=f(v^{(1)})$. If, on the other hand, $A\tilde{x}^{(0)}\leq b$ it is $f(v^{(0)})=c^\top \tilde{x}^{(0)}\leq c^\top \tilde{x}^{(1)} \leq f(v^{(1)})$.
\end{my}

\begin{myTheo} \label{Therorem_Mono_2}
Let $(A,b,c,\mathcal{L},Q)$ be a QIP and let variable $x_k$, $1\leq k \leq n$, be monotone with all non-negative entries. For any node $v$ at depth $k-1$ and its two successors $v^{(0)}$ and $v^{(1)}$ representing the assignment of $x_k=0$ and $x_k=1$, respectively, it holds: $f(v^{(0)})\leq f(v^{(1)})$.
\end{myTheo}

\begin{my}
Let there be an optimal winning strategy for the subtree of $v^{(1)}$. Due to Theorem \ref{Therorem_Mono} this strategy is also a winning strategy for the subtree of $v^{(0)}$ with all leaf values being smaller than or equal to the leaves of the strategy at $v^{(1)}$. Hence, $f(v^{(0)}) \leq f(v^{(1)})$. If there is no winning strategy for the subtree of $v^{(1)}$ it is obviously $f(v^{(0)})\leq+\infty=f(v^{(1)})$.
\end{my}

Using this easily verifiable monotonicity allows us to omit certain subtrees a priori since solving the subtree of its sibling is guaranteed to yield the desired minimax value. Obviously, similar results can be achieved for monotone variables with non-positive entries.\\

\begin{figure}[h]
\centering
\scalebox{0.75}{
\begin{tikzpicture}[level 1/.style={sibling distance = 3.5cm},
level 2/.style={sibling distance = 2cm},
level 3/.style={sibling distance = 1.6cm},
level 4/.style={sibling distance = 1cm},
level 5/.style={sibling distance = 0.8cm},
level 6/.style={sibling distance = 0.6cm},
level 7/.style={sibling distance = 0.4cm},
level 8/.style={sibling distance = 0.3cm},
level 9/.style={sibling distance = 0.2cm},
level 10/.style={sibling distance = 0.1cm},
level distance = 0.35cm] 

\node [arn_y] {}
     child{ node [arn_x] {} 
            child{ node [arn_y, xshift=-0.7cm] {}
							child{ node [arn_x] {}
            						        child{ node [arn_y, xshift=-0.35cm] {}
            						        			child{ node [arn_x] {}
            						        					child{ node [arn_y, xshift=-0.25cm] {}
									   						child{ node [arn_x] {}
									   							child{ node [arn_y, xshift=-0.15cm] {}
									   								child{ node [arn_x] {}
									   									child{ node [arn_y, xshift=-0.15cm] {}}
									   								}
									   								child{ node [arn_x] {}
									   									child{ node [arn_y, xshift=0.15cm] {}}
									   								}
									   							}
									   						}
									   						child{ node [arn_x] {}
									   							child{ node [arn_y, xshift=0.15cm] {}
									   								child{ node [arn_x] {}
									   									child{ node [arn_y, xshift=-0.15cm] {}}
									   								}
									   								child{ node [arn_x] {}
									   									child{ node [arn_y, xshift=-0.15cm] {}}
									   								}
									   							}
									   						}
									   					}
            						        			}
            						        			child{ node [arn_x] {}
            						        					child{ node [arn_y, xshift=-0.25cm] {}
									   						child{ node [arn_x] {}
									   							child{ node [arn_y, xshift=0.15cm] {}
									   								child{ node [arn_x] {}
									   									child{ node [arn_y, xshift=-0.15cm] {}}
									   								}
									   								child{ node [arn_x] {}
									   									child{ node [arn_y, xshift=0.15cm] {}}
									   								}
									   							}
									   						}
									   						child{ node [arn_x] {}
									   							child{ node [arn_y, xshift=0.15cm] {}
									   								child{ node [arn_x] {}
									   									child{ node [arn_y, xshift=0.15cm] {}}
									   								}
									   								child{ node [arn_x] {}
									   									child{ node [arn_y, xshift=0.15cm] {}}
									   								}
									   							}
									   						}
									   					}
            						        			}
            						        }
            					}                 
							child{ node [arn_x] {}
									   child{ node [arn_y, xshift=0.35cm] {}
												child{ node [arn_x] {}
									   					child{ node [arn_y, xshift=-0.25cm] {}
									   						child{ node [arn_x] {}
									   							child{ node [arn_y, xshift=-0.15cm] {}
									   								child{ node [arn_x] {}
									   									child{ node [arn_y, xshift=-0.15cm] {}}
									   								}
									   								child{ node [arn_x] {}
									   									child{ node [arn_y, xshift=-0.15cm] {}}
									   								}
									   							}
									   						}
									   						child{ node [arn_x] {}
									   							child{ node [arn_y, xshift=-0.15cm] {}
									   								child{ node [arn_x] {}
									   									child{ node [arn_y, xshift=0.15cm] (B) {}}
									   								}
									   								child{ node [arn_x] {}
									   									child{ node [arn_y, xshift=0.15cm] {}}
									   								}
									   							}
									   						}
									   					}
									   			}
            						        			child{ node [arn_x] {}
            						        					child{ node [arn_y, xshift=0.25cm] {}
									   						child{ node [arn_x] {}
									   							child{ node [arn_y, xshift=-0.15cm] {}
									   								child{ node [arn_x] {}
									   									child{ node [arn_y, xshift=-0.15cm] {}}
									   								}
									   								child{ node [arn_x] {}
									   									child{ node [arn_y, xshift=0.15cm] {}}
									   								}
									   							}
									   						}
									   						child{ node [arn_x] {}
									   							child{ node [arn_y, xshift=0.15cm] {}
									   								child{ node [arn_x] {}
									   									child{ node [arn_y, xshift=-0.15cm] {}}
									   								}
									   								child{ node [arn_x] {}
									   									child{ node [arn_y, xshift=-0.15cm] {}}
									   								}
									   							}
									   						}
									   					}
            						        			}
            						       }
            					}
            }        edge from parent node[left,,xshift=-.1cm,yshift=.15cm] {$x^\forall=0$};                     
    }
    child{ node [arn_x] {} 
            child{ node [arn_y, xshift=0.7cm] {}
							child{ node [arn_x] {}
            						        child{ node [arn_y, xshift=0.35cm] {}
            						        			child{ node [arn_x] {}
            						        					child{ node [arn_y, xshift=-0.25cm] {}
									   						child{ node [arn_x] {}
									   							child{ node [arn_y, xshift=0.25cm] {}
									   								child{ node [arn_x] {}
									   									child{ node [arn_y, xshift=-0.25cm] {}}
									   								}
									   								child{ node [arn_x] {}
									   									child{ node [arn_y, xshift=-0.25cm] {}}
									   								}
									   							}
									   						}
									   						child{ node [arn_x] {}
									   							child{ node [arn_y, xshift=0.25cm] {}
									   								child{ node [arn_x] {}
									   									child{ node [arn_y, xshift=-0.25cm] {}}
									   								}
									   								child{ node [arn_x] {}
									   									child{ node [arn_y, xshift=-0.25cm] {}}
									   								}
									   							}
									   						}
									   					}
            						        			}
            						        			child{ node [arn_x] {}
            						        					child{ node [arn_y, xshift=-0.25cm] {}
									   						child{ node [arn_x] {}
									   							child{ node [arn_y, xshift=0.25cm] {}
									   								child{ node [arn_x] {}
									   									child{ node [arn_y, xshift=-0.25cm] {}}
									   								}
									   								child{ node [arn_x] {}
									   									child{ node [arn_y, xshift=-0.25cm] {}}
									   								}
									   							}
									   						}
									   						child{ node [arn_x] {}
									   							child{ node [arn_y, xshift=0.25cm] {}
									   								child{ node [arn_x] {}
									   									child{ node [arn_y, xshift=-0.25cm] {}}
									   								}
									   								child{ node [arn_x] {}
									   									child{ node [arn_y, xshift=-0.25cm] {}}
									   								}
									   							}
									   						}
									   					}
            						        			}
            						        }
            					}                 
							child{ node [arn_x] {}
									   child{ node [arn_y, xshift=0.35cm] {}
												child{ node [arn_x] {}
									   					child{ node [arn_y, xshift=-0.25cm] {}
									   						child{ node [arn_x] {}
									   							child{ node [arn_y, xshift=0.25cm] (A)  {}
									   								child{ node [arn_x] {}
									   									child{ node [arn_y, xshift=-0.25cm] {}}
									   								}
									   								child{ node [arn_x] {}
									   									child{ node [arn_y, xshift=-0.25cm] {}}
									   								}
									   							}
									   						}
									   						child{ node [arn_x] {}
									   							child{ node [arn_y, xshift=0.25cm] {}
									   								child{ node [arn_x] {}
									   									child{ node [arn_y, xshift=-0.25cm]  {}}
									   								}
									   								child{ node [arn_x] {}
									   									child{ node [arn_y, xshift=-0.25cm] {}}
									   								}
									   							}
									   						}
									   					}
									   			}
            						        			child{ node [arn_x] {}
            						        					child{ node [arn_y, xshift=-0.25cm] {}
									   						child{ node [arn_x] {}
									   							child{ node [arn_y, xshift=0.25cm] {}
									   								child{ node [arn_x] {}
									   									child{ node [arn_y, xshift=-0.25cm] {}}
									   								}
									   								child{ node [arn_x] {}
									   									child{ node [arn_y, xshift=-0.25cm] {}}
									   								}
									   							}
									   						}
									   						child{ node [arn_x] {}
									   							child{ node [arn_y, xshift=0.25cm] {}
									   								child{ node [arn_x] {}
									   									child{ node [arn_y, xshift=-0.25cm] {}}
									   								}
									   								child{ node [arn_x] {}
									   									child{ node [arn_y, xshift=-0.25cm] {}}
									   								}
									   							}
									   						}
									   					}
            						        			}
            						       }
            					}
            }       edge from parent node[right,xshift=.1cm,yshift=.15cm] {$x^\forall=1$};                     
    }
; 
 \node[dashed, dash pattern=on 6pt off 2pt, draw, ellipse, ultra thick, minimum width=3.7cm,minimum height=.7cm, color=black,xshift=-.3cm, yshift=.17cm] at (A) {};
  \node[dotted,draw, ellipse, ultra thick, minimum width=5cm,minimum height=.7cm, color=black,xshift=-5.45cm, yshift=.17cm] at (A) {};
\end{tikzpicture}
}

\caption{Illustrative strategy for which the universal assignment $x^\forall=1$ entails a simple winning strategy:  Regardless of future universal decisions existential variables can be set in a certain simple way, e.g. the existential decisions in the dashed ellipse are all the same. $x^\forall=0$ on the other hand compels a more clever strategy, e.g. the existential decisions in the dotted ellipse differ depending on previous universal decisions.}\label{Fig_Tree}
\end{figure}
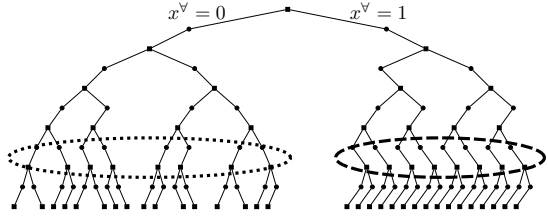

In contrast to this usage of prior knowledge we also want to gather \textit{deep knowledge}  during the search process: found strategies in certain subtrees can be useful in order to assess the minimax value of related subtrees rapidly. The idea is based upon the observation that typically in only a rather small part of the game tree a distinct and crafty strategy is required in order to ensure the fulfillment of the constraint system: in the right-hand side subtree of Figure \ref{Fig_Tree} it suffices to find a fulfilling existential variable assignment for only one scenario (universal variable assignment) and reuse it in the other branches.

\begin{myTheo}\label{Theo_CopyPaste}[Strategic Copy-Pruning (SCP)]~\\
Let $k\in \mathcal{A}$ and let $(\tilde{x}_1,\ldots,\tilde{x}_{k-1}) \in \{0,1\}^{k-1}$ be a fixed variable assignment of the variables $x_1, \ldots, x_{k-1}$. Let $v\in V_\forall$ be the corresponding node in the game tree. Let $\tilde{w} \in V$ and $\hat{w} \in V$ be the two children of $v$ corresponding to the variable assignment $\tilde{x}_k$ and $\hat{x}_k=1-\tilde{x}_k$ of the universal variable $x_k$, respectively. Let there be an optimal winning strategy for the subtree below $\tilde{w}$ with minimax value $f(\tilde{w})=\tilde{z}$ defined by the variable assignment $\tilde{x}=(\tilde{x}_1,\ldots,\tilde{x}_{n}) \in \{0,1\}^{n}$, i.e.   $\tilde{z}=c^\top \tilde{x}$. If the minimax value of the copied strategy for the subtree below $\hat{w}$ - obtained by adoption of future\footnote{\textit{future} means variable blocks with index $\geq k$.} existential variable assignments as in $\tilde{x}$ - is not larger than $\tilde{z}$ and if this copied strategy constitutes a winning strategy then $f(v)=\tilde{z}$. Formally: If both 
\begin{equation}\label{ZFBedingung2}
c_k(\hat{x}_k - \tilde{x}_k)+ \sum_{\substack{j\in\mathcal{A},\ j>k \\ \text{\normalfont and } c_j \geq 0} }  c_j(1-\tilde x_j) - \sum_{\substack{j\in\mathcal{A},\  j>k \\\text{\normalfont and } c_j < 0} } c_j \tilde x_j \leq 0
\end{equation}
and
\begin{equation}\label{CheckAllConstraints}
\sum\limits_{\substack{j \in \mathcal{E} \\ \text{\normalfont or }  j<k}}A_{i,j} \tilde{x}_j + A_{i,k}\hat x_k +\sum\limits_{\substack{ j \in \mathcal{A},\ j>k\\ \text{\normalfont and }A_{i,j}>0}} A_{i,j} \leq b_i 
\end{equation}
for all constraints $i\in \{1,\ldots,m\}$ then $f(v)=\tilde{z}$. 

\end{myTheo}

For clarification note that Condition (\ref{ZFBedingung2}) ensures that the change in the minimax value of the copied strategy, resulting from flipping $x_k$ and using the worst case assignment of the remaining future universal variables, is not positive, i.e. that its minimax value is still smaller than or equal to $\tilde{z}$. Condition (\ref{CheckAllConstraints}) verifies that every constraint is satisfied in each leaf of the copied strategy by ensuring the fulfillment of each constraint in its specific worst case scenario.

\begin{my}
If (\ref{CheckAllConstraints}) is satisfied there automatically exists a winning strategy for the subtree of $v$ corresponding to $x_k=\hat{x}_k$ with root node $\hat{w}$, since for any future universal variable assignment the assignment of upcoming existential variables as in  $\tilde{x}$ fulfills the constraint system. Further, the minimax value $\hat{z}$ of this strategy is smaller than or equal to $\tilde{z}$  due to Condition (\ref{ZFBedingung2}):
\begin{align*}
\hat{z} &= \sum\limits_{\substack{j \in \mathcal{E} \\ \text{\normalfont or }  j<k}} c_j\tilde{x}_j +c_k \hat{x}_k + \sum\limits_{\substack{j\in\mathcal{A},\ j>k \\ \text{\normalfont and } c_j \geq 0} } c_j\\
&\stackrel{ (\ref{ZFBedingung2})}{\leq}  \sum\limits_{\substack{j \in \mathcal{E} \\ \text{\normalfont or }  j<k}} c_j\tilde{x}_j +c_k \tilde{x}_k + \sum\limits_{j\in \mathcal{A},\ j>k}  c_j \tilde{x}_j \quad = \tilde{z}
\end{align*}
Hence, the (still unknown) optimal strategy for the subtree below $\hat{w}$ has a minimax value smaller than or equal to $\tilde{z}$, i.e. $f(\hat{w})\leq\hat{z} \leq \tilde{z}= f(\tilde{w})$. Therefore, with Definition \ref{Def_MinMax}, $f(v)=f (\tilde{w})=\tilde{z}$.
\end{my}

Note that, since  $A\tilde{x}\leq b$,  Condition (\ref{CheckAllConstraints}) is trivially fulfilled for any constraint $i\in \{1,\ldots,m\}$ with $A_{i,j}=0$ for all $j\in \mathcal{A}, j\geq k$, i.e. constraints  that are not influenced by future universal variables do not need  to be examined. Hence, only a limited number of constraints need to be checked in case of a sparse matrix. 
 Further, note that (\ref{ZFBedingung2}) is fulfilled if $c_j=0$ for all $j \in \mathcal{A}, j\geq k$, i.e. if the  future universal variables have no direct effect on the objective value. In particular, if $c=0$, i.e. it is a satisfiabilty problem rather than an optimization problem, Condition (\ref{ZFBedingung2}) can be neglected as it is always fulfilled.

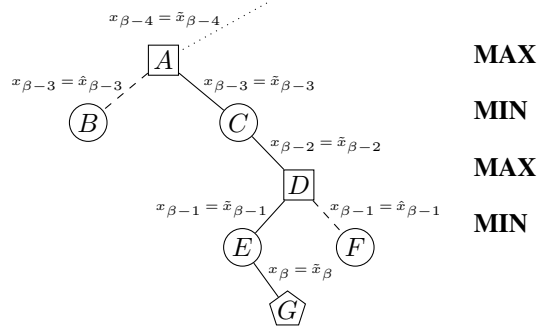
\begin{figure}[h!]
\centering
\begin{tikzpicture}[-,>=stealth',level 1/.style={sibling distance=2cm},
level 2/.style={sibling distance=2cm},
level 3/.style={sibling distance=2.5cm}, 
level 4/.style={sibling distance=1.5cm}, 
level 5/.style={sibling distance=1.7cm},
level distance=.83cm] 

\node [opacity=1]() {}
child{ node [DecisionNodeMax,xshift=-1.5cm] {$A$}
	 	child{ node [DecisionNodeMin] {$B$};\path edge from parent[dashed] node[left,yshift=.1cm,xshift=.1cm]{\tiny $x_{\beta-3}$ $=$ $\hat{x}_{\beta-3}$}
	 	}
      child{ node [DecisionNodeMin] {$C$}
      		 		child{ node [DecisionNodeMax,xshift=.8cm] {$D$}
                		child{ node [DecisionNodeMin] {$E$}
                			child{ node [Leaf,xshift=.6cm] {$G$};\path edge from parent node[right,yshift=.1cm,xshift=-.1cm]{\tiny $x_{\beta}$ $=$ $\tilde{x}_{\beta}$}
                			};\path edge from parent node[left,yshift=.1cm,xshift=.1cm]{\tiny $x_{\beta-1}$ $=$ $\tilde{x}_{\beta-1}$}
                			}
                			child{ node [DecisionNodeMin] {$F$};\path edge from parent[dashed] node[right,yshift=.1cm,xshift=-.1cm]{\tiny $x_{\beta-1}$ $=$ $\hat{x}_{\beta-1}$}
                			};\path edge from parent node[right,yshift=.1cm,xshift=-.1cm]{\tiny $x_{\beta-2}$ $=$ $\tilde{x}_{\beta-2}$}
                		};\path edge from parent node[right,yshift=.1cm,xshift=-.1cm]{\tiny $x_{\beta-3}$ $=$ $\tilde{x}_{\beta-3}$}
                	};\path edge from parent[dotted] node[left,yshift=.1cm,xshift=.1cm]{\tiny $x_{\beta-4}$ $=$ $\tilde{x}_{\beta-4}$}       
	}    		
; 
\node[anchor=west] at  (2.5, -.75) {\textbf{MAX}};
\node[anchor=west] at  (2.5,-1.5) {\textbf{MIN}};
\node[anchor=west] at  (2.5,-2.25) {\textbf{MAX}};
\node[anchor=west] at  (2.5,-3) {\textbf{MIN}};
\end{tikzpicture}

\caption{Illustrative game tree: Circular nodes are existential decision nodes, rectangular nodes are universal decision nodes and pentagonal nodes are leaves. The dashed lines indicate that those underlying subtrees might be omitted if Theorem \ref{Theo_CopyPaste} applies.}\label{Fig_Explanation}
\end{figure}

The theoretical result from Theorem \ref{Theo_CopyPaste} must be implemented cautiously. For a brief explanation of Algorithm \ref{Algo}  consider Figure \ref{Fig_Explanation} representing the final four variables of a QIP with strictly alternating quantifiers.

 We assume the search has found the fulfilling variable assignment $x=\tilde{x}$ (represented by node G) for which $x_{\beta}=\tilde{x}_{\beta}$ is the optimal assignment for the final variable block with regard to $x_1=\tilde{x}_1,\ldots, x_{\beta-1}=\tilde{x}_{\beta-1}$, i.e. $f(E)=f(G)$. If the requirements of Theorem \ref{Theo_CopyPaste} for $k=\beta-1$ are fulfilled it is $f(D)=f(E)$ and we do not have to calculate $f(F)$ explicitly as the existence of a winning strategy below $F$ is ensured. If this attempt is successful the application of Theorem \ref{Theo_CopyPaste} at node $A$ would be attractive. However, one must ensure, that $f(C)=f(D)$, i.e. that setting $x_{\beta-2}=\tilde{x}_{\beta-2}$ is indeed optimal in this stage. If this optimality cannot be guaranteed, but Conditions (\ref{ZFBedingung2}) and (\ref{CheckAllConstraints}) are fulfilled at node $A$, we still can conclude the existence of a winning strategy for the subtree at $B$ but we cannot yet specify $f(A)$. 
However, storing the information $f(B)\leq \hat{z}$ and $f(A)\leq \tilde{z}$ can be advantageous (see line \ref{Line_Update} in Algorithm \ref{Algo}).

\begin{algorithm}[h!]
  \SetAlgoLined
  \KwData{$\tilde{x}$, $\tilde{z}$}
  $v=$ last universal node\;
  $k=$ index of the variable associated with $v$\;
  mode=Pruning\;
    \While{$v \neq root$}{
  \If{$x_k$ is monotone and $\tilde{x}_k$ is set accordingly}{
 	mark $v$ as finished;\ goto line \ref{Line}\;
  }
  \eIf(\tcp*[h]{$x_k$ is existential variable}){$v\in V_\exists$}{
  \If{$f(v)\neq\tilde{z}$ \normalfont{ \textbf{or}} $f(v)$ unknown \label{Line_OptWinStrat}}{
  	mode=BoundUpdate;\ goto line \ref{Line}\;
  }
  }(\tcp*[h]{$x_k$ is universal variable})
  {
  \lIf{Condition (\ref{ZFBedingung2}) is violated\label{Line_Cond1}}{\textbf{return}}
  \For{each constraint $i$ with $A_{i,k}\neq 0$ \label{Line_Only}}{
  \lIf{Condition (\ref{CheckAllConstraints}) is violated for $i$ \label{Line_Cond2}}{\textbf{return}}
  }
  \eIf{mode$=$Pruning}{mark $v$ as finished\;}
  {update bound: $f(v)\leq \tilde{z}$\;\label{Line_Update}} 
  }
  $v$=predecessor($v$)\;\label{Line}
$k$=index of the variable associated with $v$\; 
  }
\caption{RecycleStrategy($\tilde{x}$, $\tilde{z}$)}\label{Algo}
\end{algorithm}  

Note, that as soon as the first universal node is found during this backtracking for which Conditions (\ref{ZFBedingung2}) or (\ref{CheckAllConstraints}) are violated Algorithm \ref{Algo} stops. \textit{Further, note that if Condition (\ref{CheckAllConstraints}) is fulfilled at some universal node, e.g. node $D$ in Figure \ref{Fig_Explanation}, for the next universal node above, e.g. node $A$, Condition (\ref{CheckAllConstraints}) only needs to be checked for those constraints $i \in \{1,\ldots,m\}$ in which the variable corresponding to the node of interest, e.g. $x_{\beta-3}$, is present (see line \ref{Line_Only} in Algorithm \ref{Algo}).} This allows a very fast verification of Condition (\ref{CheckAllConstraints}) if matrix $A$ is sparse, making Theorem \ref{Theo_CopyPaste} practically applicable.
An outline of our implementation  is given in Algorithm \ref{Algo} that either prunes subtrees, i.e. marks nodes as finished (“mode=Pruning” as long as the “optimal winning strategy”-condition of Theorem 3 is met, see line \ref{Line_OptWinStrat}), or updates the bounds on the minimax value of universal nodes (“mode=BoundUpdate” as soon as Theorem 3 cannot be applied anymore). The presented function is invoked, when for fixed variables $x^{(1)}=\tilde{x}^{(1)}, \ldots, x^{(\beta-1)}=\tilde x^{(\beta-1)}$ the optimal assignment of $x^{(\beta)}=\tilde{x}^{(\beta)}$ was found as exemplarily described above. The variable allocation $x=\tilde{x}$ and the corresponding objective value $\tilde{z}=c^\top \tilde{x}$ are the input for Algorithm \ref{Algo}.

Note, that computing line \ref{Line_Cond1} of Algorithm \ref{Algo}, i.e. checking Condition \eqref{ZFBedingung2}, requires $\mathcal{O}(n-k)$ operations, while line \ref{Line_Cond2} is called $m_k$ times (with $m_k$ being the number of constraints in which the current universal variable $k$ occurs) and the computing time of line  \ref{Line_Cond2} itself is $\mathcal{O}(n-k)$. Hence, Algorithm \ref{Algo} has an overall runtime of $\mathcal{O}(m_k(n-k))$. Thus, the complexity is linear in the input size (size of matrix A). In our experiments, where each of the universal variables occurs in only a few rows and the matrix is sparse, the runtime of the heuristic is negligible. 
\begin{myex}
Let us consider the following QIP with binary variables (The min/max alternation in the objective and the binary variable domains  are omitted):
$$
\begin{array}{lrrrrrl}
\min 			& 2x_1 	& +	3x_2  	&-	2x_3	&-2x_4&+x_5	\\
\text{s.t.} 	&\exists x_1  &\forall x_2&\exists x_3&\forall x_4&\exists x_5&:	
\end{array}
$$
$$
\left( \begin{array}{rrrrr}
1 & -1 &\color{white}{+}\color{black} 1 & \color{white}{+}\color{black} 3 & -1\\
3 & 2 & 3 & 1 & -2
\end{array} \right) x \leq \left( \begin{matrix}
2\\
1
\end{matrix} \right) $$
Starting at the root node of the corresponding game tree we can immediately omit the subtree corresponding to $x_1=1$ due to the monotonicity of $x_1$.  Keep in mind that the result of Theorem \ref{Theo_CopyPaste} is particularly beneficial if the search process of a QIP solver first examines the principal variation, i.e. the variable assignment defining the actual minimax value.
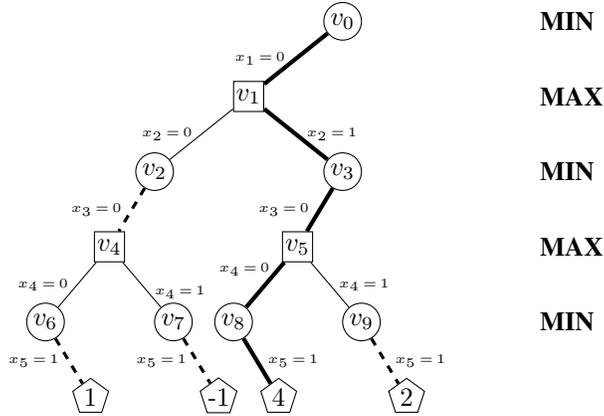
\begin{figure}[h]
\centering
\begin{tikzpicture}[-,>=stealth',level 1/.style={sibling distance=2cm},
level 2/.style={sibling distance=2.5cm}, 
level 3/.style={sibling distance=1.5cm}, 
level 4/.style={sibling distance=1.7cm},
level distance=1cm] 

\node [DecisionNodeMin](Top) {$v_0$}
	 	child{ node [DecisionNodeMax,xshift=-1.25cm] {$v_1$}
	 		child{ node [DecisionNodeMin] {$v_2$}
                	child{ node [DecisionNodeMax,xshift=-.6cm] {$v_4$}
                		child{ node [DecisionNodeMin] {$v_6$}
                			child{ node [Leaf,xshift=.6cm] {$1$};\path edge from parent[dashed, very thick] node[left]{\tiny $x_5$ $=$ $1$}	
                			};\path edge from parent[] node[left]{\tiny $x_4$ $=$ $0$}	
                		}
                		child{ node [DecisionNodeMin] {$v_7$}
                			child{ node [Leaf,xshift=.6cm] {-$1$};\path edge from parent[dashed, very thick] node[left]{\tiny $x_5$ $=$ $1$}	
                			};\path edge from parent[] node[right,yshift=-.1cm,xshift=.05cm]{\tiny $x_4$ $=$ $1$}	
                		};\path edge from parent[dashed, very thick] node[left]{\tiny $x_3$ $=$ $0$}	
                	}   ;\path edge from parent[] node[left]{\tiny $x_2$ $=$ $0$}	
          	}
          	child{ node [DecisionNodeMin] {$v_3$}
                	child{ node [DecisionNodeMax,xshift=-.6cm] {$v_5$}
                		child{ node [DecisionNodeMin] {$v_8$}
                			child{ node [Leaf,xshift=.6cm] {$4$};\path edge from parent[ultra thick] node[right]{\tiny $x_5$ $=$ $1$}	
                			};\path edge from parent[ultra thick] node[left,yshift=.2cm,xshift=.2cm]{\tiny $x_4$ $=$ $0$}	
                		}
                		child{ node [DecisionNodeMin] {$v_9$}
                			child{ node [Leaf,xshift=.6cm] {$2$};\path edge from parent[dashed, very thick] node[right]{\tiny $x_5$ $=$ $1$}	
                			};\path edge from parent[] node[right]{\tiny $x_4$ $=$ $1$}	
                		} ;\path edge from parent[ultra thick] node[left]{\tiny $x_3$ $=$ $0$}	
                	}   	 ;\path edge from parent[ultra thick] node[right]{\tiny $x_2$ $=$ $1$};
          	} ;\path edge from parent[ultra thick] node[left]{\tiny $x_1$ $=$ $0$};
       	}
; 
\node[anchor=west] at  (2.5, 0) {\textbf{MIN}};
\node[anchor=west] at  (2.5,-1) {\textbf{MAX}};
\node[anchor=west] at  (2.5,-2) {\textbf{MIN}};
\node[anchor=west] at  (2.5,-3) {\textbf{MAX}};
\node[anchor=west] at  (2.5,-4) {\textbf{MIN}};
\end{tikzpicture}
\caption{Optimal winning strategy for the stated QIP. Circular nodes are existential decision nodes (from $V_\exists$), rectangular nodes are universal decision nodes and pentagonal nodes are leaves. The values given in the leaves constitute the objective value corresponding to the variable assignment along the path from the root to this leaf. The dashed lines indicate that those existential decisions where simply copied from the path drawn thicker.} \label{Example_Tree}
\end{figure}
Assume the search process follows the path drawn thick in Figure \ref{Example_Tree} to node $v_8$, i.e. the path corresponding to the variable assignment $x_1=0$, $x_2=1$, $x_3=0$ and $x_4=0$. Setting $x_5=1$ is optimal in this case, as $x_5=0$ would violate the second constraint. Hence, the minimax value of $v_8$ is $4$. On the way up in the search tree we then want to determine $f(v_5)$. As (\ref{ZFBedingung2}) and (\ref{CheckAllConstraints}) are fulfilled for $k=4$, $\tilde{z}=4$ and $\tilde{x}=(0,1,0,0,1)$ we know that $f(v_5)=4$. That means we have (easily) verified a winning strategy starting from $v_9$ with minimax value smaller than or equal to $4$.  In node $v_3$ setting $x_3=1$ is obviously to the detriment of the existential player, because the second constraint would become unfulfillable. Hence, $f(v_3)=f(v_5)=4$. 
In node $v_1$ we once again try to apply Theorem \ref{Theo_CopyPaste} by copying the existential decisions of $x_3$ and $x_5$ in the thick path to the not yet investigated subtree associated with $x_2=0$. As (\ref{ZFBedingung2}) and (\ref{CheckAllConstraints}) are fulfilled for $k=2$, $\tilde{z}=4$ and $\tilde{x}=(0,1,0,0,1)$ this attempt is successful and $f(v_1)=4$. Note that by applying Theorem \ref{Theo_CopyPaste}  the minimax value of the subtrees below $v_2$ and $v_9$ are not known exactly: in particular we only obtain $f(v_2)\leq \hat{z}=1$, whereas a better strategy exists resulting in $f(v_2)=0$ (Setting $x_5=0$ in node $v_6$). \\
Hence, by  finding the principal variation first (thick path) and applying Theorem \ref{Therorem_Mono_2} at node $v_0$,  Theorem \ref{Theo_CopyPaste} at node $v_1$ and $v_5$ and some further reasoning from linear programming at node $v_3$ and $v_8$ the minimax value at the root node $v_0$ was found to be 4 with optimal first stage solution  $x_1=0$. 
\end{myex}

Theorem \ref{Theo_CopyPaste} can particularly  come into effect if the branching decisions at universal nodes result in rather vicious scenarios, i.e. in variable assignments restricting the constraint system and maximizing the objective value. Hence, the applicability of the presented results largely depends on the implemented diving and sorting heuristic. 
\section{Solver, Experiments and Results}\label{Sec_Exp}

The open source\footnote{We accessed the open sources from \url{http://www.q-mip.org}} solver Yasol  \cite{YasolACG17}, which is used to analyze the theoretical findings, combines two well known search mechanisms: The alpha-beta algorithm \cite{KNUTH1975293}, traditionally used in a game tree search environment, and a generalization of the DPLL algorithm \cite{Zhang}, used to solve SAT problems. We extended the main search algorithm to a scout algorithm \cite{Reinefeld}. The solver proceeds in two phases in order to find optimal solutions of 0/1-QIP instances.
\begin{itemize}
\item Phase 1 (Feasibility Phase): The instance's feasibility is determined, i.e. it is checked whether the instance has any solution at all. During this phase, the solver acts like a QBF solver \cite{Cadoli2002,Zhang} with some extra abilities. Technically it performs a null window search \cite{Pearl80}.
\item  Phase 2 (Optimization Phase): The solution space is explored in order to find the provable optimal solution. The (nega)scout algorithm  is enhanced by non-chronological backtracking and backward implication \cite{Zhang,Qube}.
\end{itemize}

We enhanced this solver in two different ways:
\begin{itemize}
\item[1.] The detection of monotone variables (MONO) was implemented and their properties exploited during the game tree search.
\item[2.] The adoption of existing winning strategies (strategic copy-pruning (SCP)) from one branch of a universal node to  another was realized.
\end{itemize} 
The SCP-enhancement (made possible by Theorem \ref{Theo_CopyPaste}) can be switched on and off in  both phases separately.

The instances used to study the effect of the presented results are runway scheduling problems under uncertainty modeled as QIPs. They were created following the ideas presented in \cite{HEIDT201628}. The task is to find a b-matching: all airplanes must be assigned to exactly one time slot, while one time slot can take in at most $b$ airplanes. Furthermore, the airplanes must land within an uncertain time windows (a set of time slots). Reasons for such variations (in the arrival time) might be adjusted airspeed (due to weather) or operational problems. Hence, we are interested in an initial matching plan that can be fixed cheaply if the mandatory time windows for some planes do not contain the initially scheduled time slot. The testset contains 29 instances\footnote{The studied benchmark instances and a brief explanation can be found at \url{http://www.q-mip.org/index.php?id=41}}, varying in the number of planes, the number of time slots, the type of allowed disturbances, the number of universal blocks and the cost function. In terms of the sizes of the (\textit{solved feasible}) instances this results in between 100-300 existential variables, 10-30 universal variables and 50-100 constraints. 

In Table \ref{Table_Results} the number of solved instances is displayed for different settings. For each instance a maximum of one hour solution time was provided.
All experiments were executed on a PC with an Intel i7-4790 (3.6 GHz) processor and 32GB RAM. 
\begin{table}[h]
\centering
\caption{Number of solved instances dependend on the solver setting: exploitation of monotone variables (MONO) and the strategic copy-pruning (SCP) in different phases of the solver.}\label{Table_Results}
\begin{tabular}{||l|l||c|}
\cline{1-2}
\multicolumn{2}{||c||}{Setting}\\\hline
MONO&SCP
 &\# solved\\ \hline \hline
off & off&14\\ \hline
off &only feas&16\\ \hline
off &only opt&21\\ \hline
off &both&24\\ \hline\hline
on & off&23\\ \hline
on &only feas&24\\ \hline
on &only opt&25\\ \hline
on &both&25\\ \hline
\end{tabular}
\end{table}

If neither of the presented procedures is used 14 out of 29 instances are solved. Without taking advantage of the monotonicity SCP can be benficial in either solution phase regarding the number of solved instances. If applied in both phases the number of solved instances is increased up to 24. When also exploiting the monotonicity the number of solved instances increases to 25. However, SCP turns out to be somewhat disadvantegous in the feasibility phase. Even though an additional instance is solved (24) compared to the setting with SCP turned off (23) the average solution time increases: in Table \ref{Table_TIME} the average time needed for the 23 instances solved by all versions with turned on monotonicity is displayed.
\begin{table}[h]
\centering
\caption{Average time needed for the 23 solved instances by all four settings and activated monotonicity. }\label{Table_TIME}
\begin{tabular}{|l|c|c|c|c|}
\hline
SCP setting&off&only feas & only opt &both\\\hline
avg. time&84.17s&101.70s&	25s& 32s\\\hline

\end{tabular}
\end{table}
Four instances were not solved at all. These instances have more than 100 universal variables and more than 10000 existential variables. However, there also are infeasible instances of the same magnitude that are solved within seconds. Nonetheless, detecting a contradiction leading to infeasibilty can obviously be much faster than finding and ensuring an optimal strategy with more than $2^{100}$ leaves.

The best setting is to use SCP only in the optimization phase while exploiting variable monotonicity because additionally using SCP in the feasibility phase slightly increases the average solution time. Our conjecture is that this is due to biasing effects. Experiments conducted on a QBF test collection\footnote{QBF instances can easily be converted into the QIP format.} of 797 instances, taken from \url{www.qbflib.org}, show positive effects for the SCP version. With the setting 'Mono on' and 'SCP off' 644 instances are solved. If SCP is turned on in both phases (it actually is only invoked in the feasibility phase as no optimization phase is conducted) 674 instances can be solved. Further, the solution time on the instances solved in both cases decreased by 15\% when SCP is used.

In order to assess the performance results, we also built the deterministic equivalent program of each instance of the 29 runway scheduling instances and tried to solve the resulting integer program using CPLEX 12.6.1.0, a standard MIP solver. Only six of the 29 instances where solved this way, given the same amount of time (one hour), while for 14 instances not even the construction of the corresponding DEP could be finished, some of them because of the limited memory of 32 GB RAM.
 
\section{Conclusion}\label{Sec_Con}
We introduced the concept of strategic copy-pruning (SCP) during game tree search for quantified integer programs, which are robust multistage optimization problems. SCP makes it possible to omit certain subtrees during the game tree  search by implicitly verifying the existence of a strategy in linear time: finding a single leaf and applying SCP can be sufficient to guarantee an optimal strategy in a subtree. This is standing in contrast to existing algorithms such as Kawano's simulation, (nega)scout, sss* and MTD(f) in which the existence of a strategy is proven by traversing it explicitly. In addition to the theoretical results, we presented how those findings can be applied in a game tree search environment. Even though the generalized theoretical result implies linear computing time in the number of non-zero-elements of the constraint matrix, the presented partial realization of SCP as well as the sparsity of matrices allow high-speed pruning. Experiments showed that utilizing the presented approach in the open source solver Yasol resulted in a massive boost in both the number of solved instances and the solution time on a particular testset. Because of the strictly formal framework provided by QIPs we were able to derive the SCP procedure. It would be interesting to see whether SCP can be transferred to other areas of optimization under uncertainty.

\fontsize{9.0pt}{10.0pt} \selectfont
\bibliography{AAAI2019}
\bibliographystyle{abbrv}

\end{document}